\documentclass{article}
\pdfoutput=1
\usepackage{multicol}
\usepackage{lipsum}

\usepackage{graphicx}
\usepackage{subfig}
\usepackage{amsmath}
\usepackage{epstopdf}

\title{An MDM Spin Transparent Quadrupole for Storage Ring based EDM Search}
\author{M. Bai, Y. Dutheil\\
Forschungszentrum, Juelich, Germany\\
D. Sagan\\
Cornell University, Ithaca, NY, USA}

\begin{document}

\maketitle

\begin{abstract}

A storage ring provides an attractive option for directly measuring the electric dipole moment~(EDM) of charged particles. To reach a sensitivity of $10^{-29}$~e$\cdot$cm, it is critical to mitigate the systematic errors from all sources. This daunting task is pushing the precision frontier of accelerator science and technology beyond its current state of the art. Here, we present a unique idea of a magnetic dipole moment~(MDM) spin transparent quadrupole that can significantly reduce the systematic errors due to the transverse electric and magnetic fields that particle encounters.
\end{abstract}

\section{Introduction}
The quest for understanding the fate of antimatter in the universe -- and to fully understand CP violation -- has triggered a hunt to measure the electric dipole moment~(EDM). So far, all measurements have shown the upper limit of the neutron EDM is on the order of $10^{-26}$~e$\cdot$cm. However, no direct EDM measurement of charged light ions including protons, have yet been performed. The concept of a storage ring based EDM search has been proposed and investigated by the Brookhaven National Laboratory~(BNL) as well as by the Forschungszentrum Juelich~(FZJ)~\cite{srEDM,JEDI}.

In a storage ring, the spin motion is governed by the Thomas-BMT equation
\begin{equation}
\small \frac{d\vec{S}}{dt} =   \vec{S}\times \frac{e}{m}       ([(\frac{1}{\gamma}+G) \vec{B}_{\bot}+  (\frac{1+G}{\gamma})\vec{B}_{\parallel}+(G+\frac{1}{\gamma+1})\frac{\vec{E}\times \vec{\beta}}{c}] +[\frac{\eta }{2c}(\vec{E}+{c\vec{\beta}\times\vec{B}})])
\label{bmt}
\end{equation}
where $\vec{E}$ and $\vec{B}$ are the electric and magnetic fields in the laboratory frame, $G$ is the anomalous g-factor of the particle, $\vec{d}=\frac{e \eta}{2m}\vec{S}$ is the electric dipole moment, $\vec{\mu}=\frac{g}{2}\frac{e}{m}\vec{S}$ is the magnetic dipole moment, and $\vec{\beta}$ is the particle's velocity normalized by the speed of the light. In Eq.~\ref{bmt}, $\vec{S}$ is the spin vector in the particle's frame, while $\vec{B}_{\bot}$ and $\vec{B}_{\parallel}$ are the magnetic field perpendicular and parallel to the particle's velocity, respectively.

Since the EDM is expected to be significantly smaller than the magnetic dipole moment~(MDM), the spin motion is dominated by the MDM. The concept of measuring the EDM in a storage ring with the MDM part of the spin motion frozen along the velocity of the particle was developed and investigated by the BNL srEDM collaboration\cite{dEDM_BNL,pEDM_BNL}. The proposed storage ring for a deuteron EDM search uses static hybrid electro and magnetic deflectors to freeze the MDM part of the spin motion. The possible sources of systematic errors for such a high precision EDM storage ring was evaluated and the residual non-zero average vertical electric field was identified as the only source of first order systematic errors. Other error sources, such as fringe fields as well as the electric and magnetic fields in the RF cavity, introduce higher order systematic errors to the EDM measurement. The spin precession due to the average non-zero vertical electric field $\left\langle E_v\right\rangle$ is given by\cite{dEDM_BNL}
\begin{equation}
\omega_{E_v}=(G+1)\frac{e}{mc}\frac{\left\langle E_v \right\rangle}{\beta^2\gamma^2}.
\label{Ev_BNL}
\end{equation}

For a partially spin frozen based EDM search in a storage ring, such as the proposed direct measurement of the deuteron EDM at the Cooler SYnchrotron~(COSY)~\cite{JEDI,yannis}, the additional vertical polarization buildup from the radial magnetic fields is the dominant source of systematic error. Hence, the closed orbit has to be corrected and controlled to a high precision~\cite{marcel}. This is currently the limiting factor for achieving the proposed measurement sensitivity. Hence, the MDM spin transparent quadrupole proposed below may also be beneficial by mitigating the systematic error from the radial magnetic fields due to the off-center trajectory in the magnetic quadrupole fields. 

\section{MDM Spin Transparent Quadrupole}
Similar to the condition for freezing spin in a dipole, in order to make the MDM part of the spin motion to be transparent in a quadrupole, both electric and magnetic quadrupole fields are needed. For a linear electric quadrupole, the electric field can be described as
\begin{equation}
E_x-iE_y=(b_{e1}+ia_{e1})\frac{x+iy}{r_0}
\end{equation}
with $r_0$ being the radius of the quadrupole, $b_{e1}$ and $a_{e1}$ being the normal and skew component, respectively. $x$ and $y$ are the horizontal and vertical distances from the center of the quadrupole. The magnetic field of a quadrupole can be described as
\begin{equation}
B_y+iB_x=(b_{1}+ia_{1})(x+iy)
\end{equation}
with $b_{1}$ and $a_{1}$ being the magnetic normal component and skew component of the quadrupole, respectively.

Two conditions have to be met to achieve spin transparency. First of all, the electric and magnetic fields have to be perpendicular to each other everywhere in the quadrupole. If there are no electric nor magnetic skew fields, that is, $a_{e1} = a_1 = 0$, it is easily shown that the electric and magnetic fields stay perpendicular everywhere if
\begin{equation}
\vec{B} \cdot \vec{E}=0.
\end{equation}

Similar to the approach for spin frozen deflector~\cite{Ed}, the strength of the electric and magnetic fields for the MDM spin transparent quadrupole have to follow the given relationship in Eq.~\ref{spintransparent_BE_relation},
\begin{eqnarray}
\nonumber
B_x= - \left[ 1-\frac{1}{(\gamma +1)(1+G\gamma)} \right] \frac{\beta E_y}{c} \\
B_y= \left[ 1-\frac{1}{(\gamma +1)(1+G\gamma)} \right] \frac{\beta E_x}{c}.
\label{spintransparent_BE_relation}
\end{eqnarray}	
Thus,
\begin{equation}
b_1= \left[ 1-\frac{1}{(\gamma +1)(1+G\gamma)} \right] \frac{\beta b_{e1}}{r_0c}. 
\end{equation}

The equivalent field gradient of a combined EB quadrupole is $k_1=b_1+\frac{b_{e1}}{r_0\beta c}$. Thus, the equivalent field gradient of an MDM spin transparent quadrupole is
\begin{equation}
k_1= \left[ \left[ 1-\frac{1}{(\gamma +1)(1+G\gamma)} \right] \frac{\beta}{c}+\frac{1}{\beta c} \right] \frac{b_{e1}}{r_0}.
\end{equation}
For a 1.0~GeV/c deuteron in a combined quadrupole with a field gradient of $k_1=1.0$~$m^{-1}$, the corresponding electric field gradient is 6.4~MV/m at a radius of 50~mm and the magnetic field gradient is 0.2~T/m.

\section{Benefits of MDM Spin Transparent Quadrupole}
As discussed in the original BNL deuteron EDM storage ring proposal, where the lattice consists of spin frozen electric and magnetic deflectors, the residual average non-zero vertical electric field $\left\langle E_v \right\rangle$ is the primary source of first order contribution to the systematic error. In the BNL proposal~\cite{dEDM_BNL}, the additional spin precession due to the average vertical electric field is given by Eq.~\ref{Ev_BNL}.

For deuterons with a momentum of $1$~$GeV/c$, $G=-0.14$ and $\gamma=1.14$, the additional spin precession due to non-zero vertical electric field is about $2.0\frac{e}{mc} \left\langle E_v \right\rangle$.

If MDM spin transparent (MDMST) quadrupoles are added to the lattice, the non-zero vertical electric field will force the particles to experience additional Lorentz forces from all the quadrupoles to stay stable. In this case the average of electric and magnetic forces due to the quadrupoles will balance the residual electric field.
\begin{equation}
\left\langle \beta c B_{qx} + E_{qy} \right\rangle+\left\langle E_v\right\rangle = 0
\end{equation}
where $B_{qx}$ and $E_{qy}$ are the radial magnetic field and vertical electric field a particle sees in a MDMST quadrupole, and the resulting spin precession $\omega_{\text{MDMST}}=0$. So, the spin precession due to non-zero average vertical electric field is given by Eq.~\ref{EB_omegaEv}
\begin{equation}
\omega_{E_v}=\frac{e}{mc}(G+\frac{1}{\gamma+1})\left\langle E_v\right\rangle \beta.
\label{EB_omegaEv}
\end{equation}
For deuterons at momentum of $1$~$GeV/c$, $G=-0.14$ and $\gamma = 1.14$, this additional spin precession due to non-zero vertical electric field is about $0.2\frac{e}{mc} \left\langle E_v \right\rangle$, about an order of magnitude smaller than the case with pure magnetic quadrupoles.

\subsection{Contribution to the EDM Signal}
Using combined electric and magnetic quadrupoles, a non-zero closed orbit results to a net contribution to the EDM part of the spin precession of
\begin{equation}
\frac{d\vec{S}}{dt}=\frac{e}{\gamma m}\frac{\eta}{2c}\vec{S}\times [E_{bx}+c\vec{\beta}\times B_{by}+\frac{b_{e1}}{r_0}(x\hat{x}-y\hat{y}) +c\vec{\beta}\times b_1 (x\hat{x}+y\hat{y})]
\label{EBringEDMbuildup}
\end{equation}
Where $E_{bx}$ and $B_{by}$ are the radial electric and vertical magnetic fields of the deflector, and $c\vec{\beta}$ is the particle's velocity. $b_1$ and $b_{e1}$ are the magnetic and electric field gradients of the quadrupole, and $x$ and $y$ are the radial and vertical closed orbit distortions in the quadrupole. Unlike a pure electric or magnetic deflector, this perturbation on the EDM part of the spin motion depends on the beam position as shown in Eq.~\ref{EBringEDMbuildup}

Again, as long as beam motion is stable, the average net Lorentz force from the quadrupoles has to be zero on the closed orbit, That is, 
\begin{equation}
\left\langle \frac{b_{e1}}{r_0}(x\hat{x}-y\hat{y}) +c\vec{\beta}\times b_1 (x\hat{x}+y\hat{y}) \right\rangle = 0.
\end{equation} 
Hence, the average EDM buildup from the MDMST quadrupoles has to be also averaged to zero, and the net EDM buildup rate is given by
\begin{equation}
\left\langle \frac{d\vec{S}}{dt}\right\rangle=\frac{e}{\gamma m}\frac{\eta}{2c}\vec{S}\times [E_{bx}+c\vec{\beta}\times B_{by}].
\label{EBringEDMavg}
\end{equation}

\break

\begin{figure}[bt]
	\subfloat[ring layout]{\includegraphics[width=0.40\textwidth]{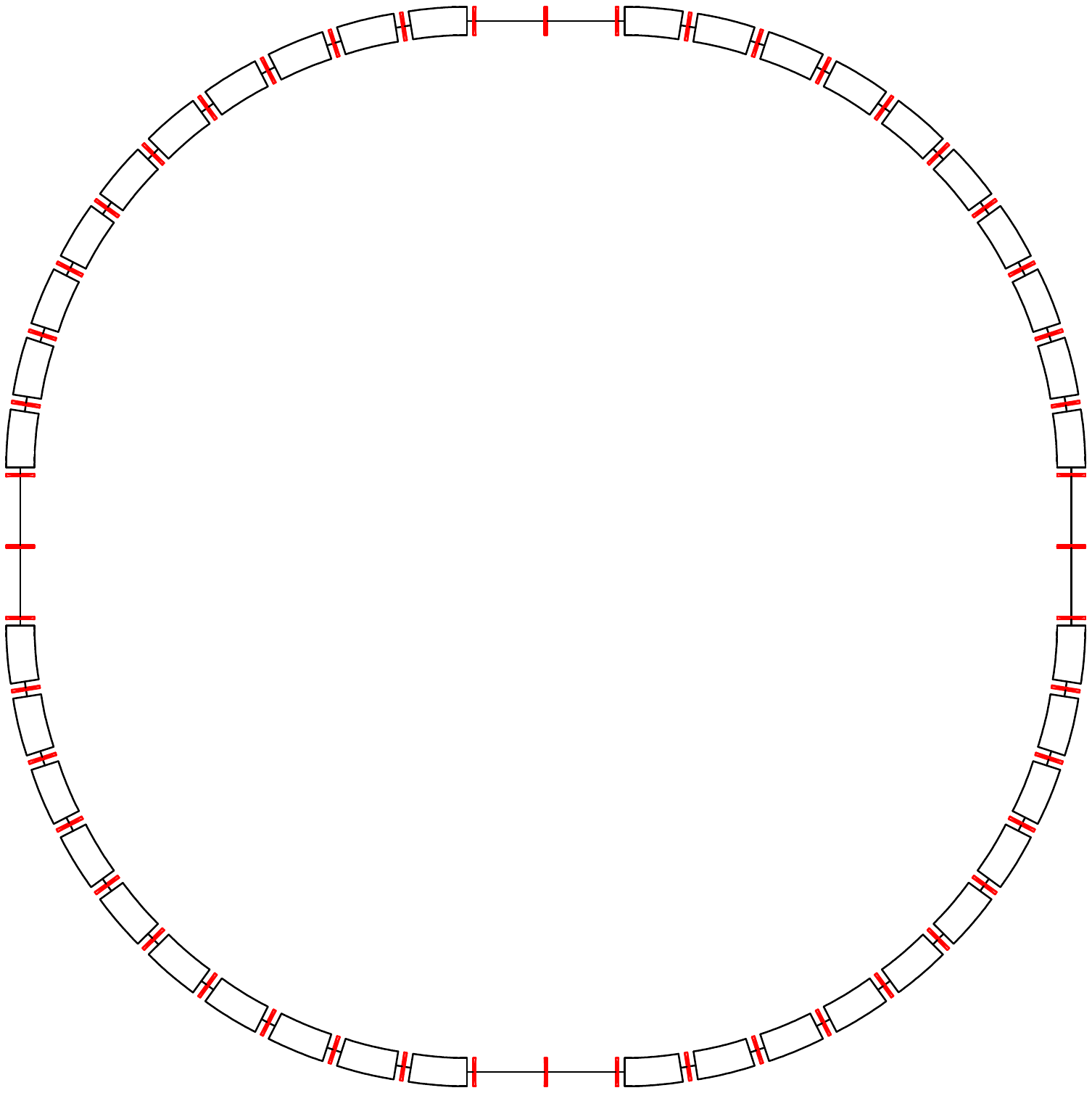}\label{fig:layout}}
	\hfill
	\subfloat[Twiss and dispersion]{\includegraphics[width=0.48\textwidth]{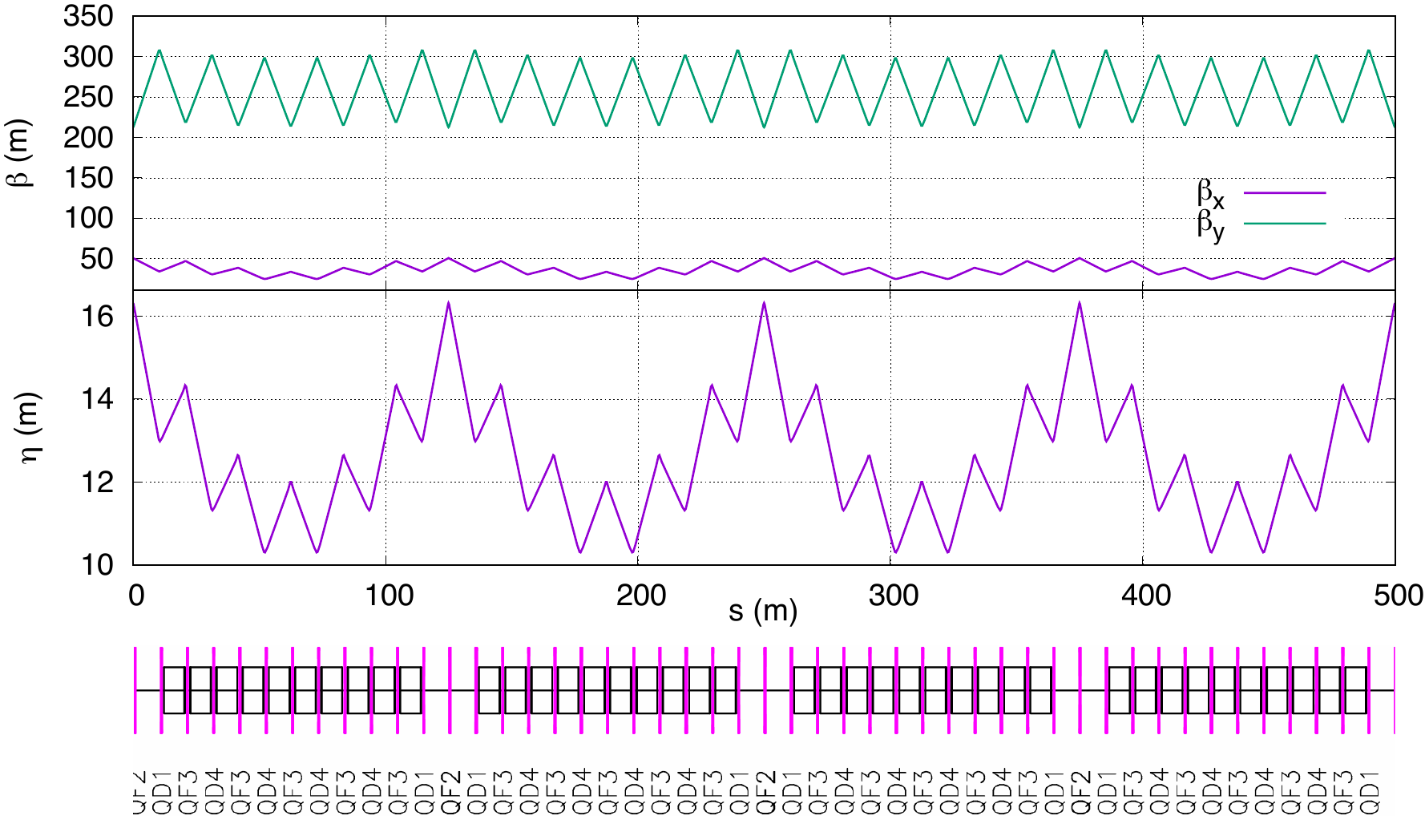}\label{fig:twiss}}
	\caption{schematic layout of deuteron EDM storage ring with MDMST quadrupoles and full spin frozen EB deflectors, The top and middle plot of Fig.~\ref{fig:twiss} shows the beta functions and dispersion function, respectively. The horizontal and vertical betatron tunes are 2.31 and 0.312.}
	\label{ringlayout}
\end{figure}

\begin{figure}[bt]
	\subfloat[deuteron EDM  ring ]{\includegraphics[width=0.48\textwidth]{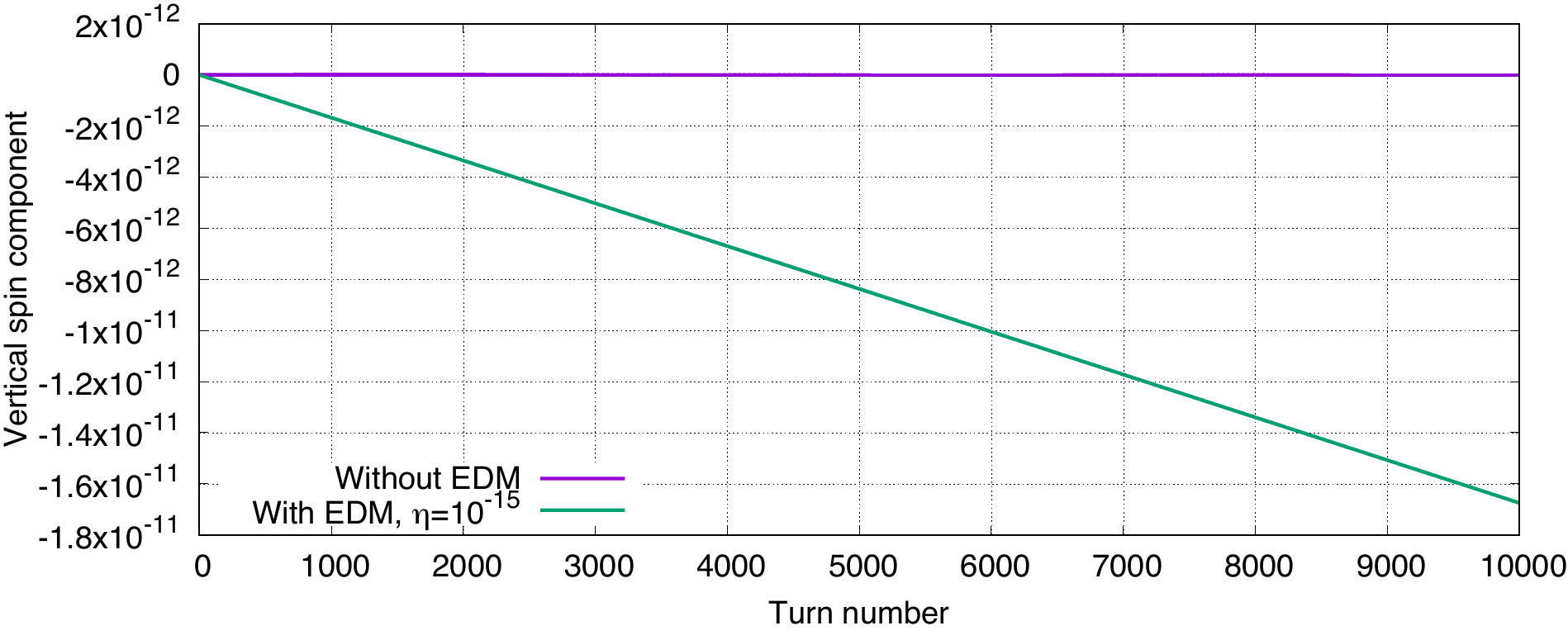}}
  \hfill
  \subfloat[proton EDM ring]{\includegraphics[width=0.48\textwidth]{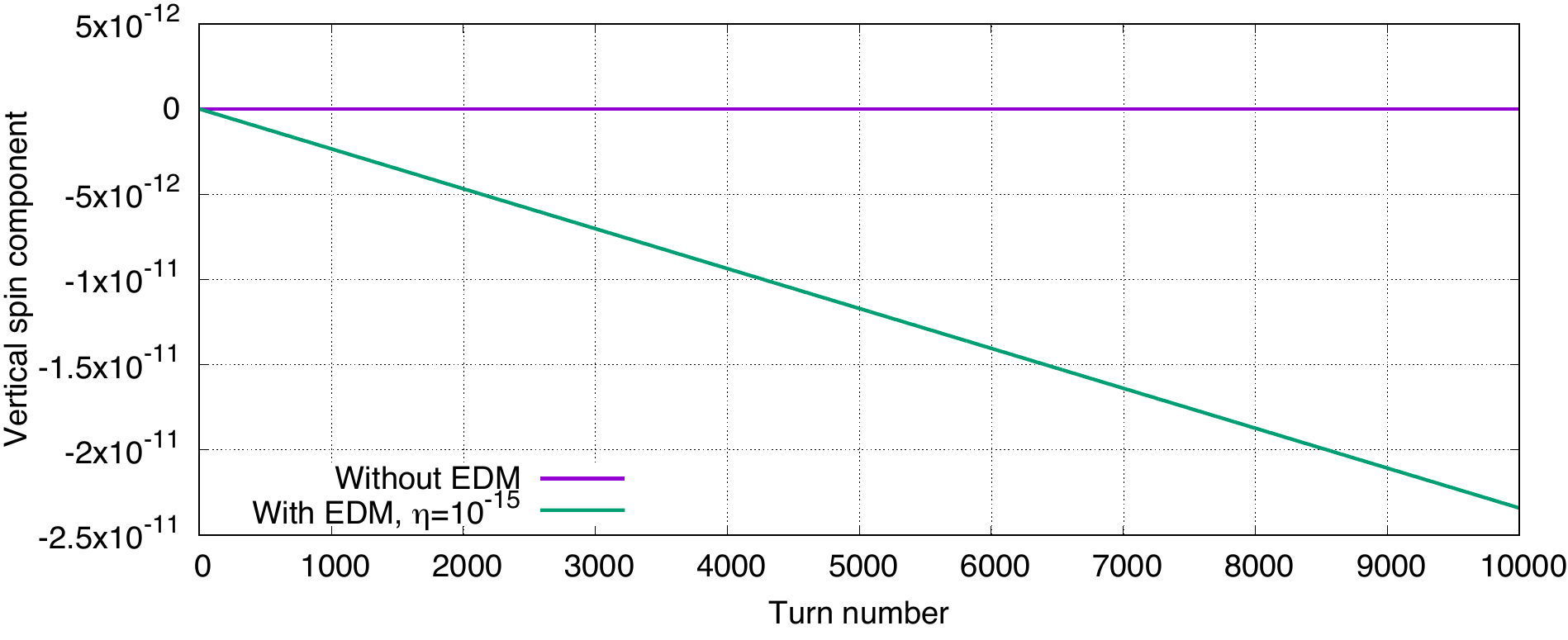}}
	\caption{Average vertical spin component of 8 particles as a function of turn number. a) Using a deuteron ring with MDMST quadrupoles. b) Using a proton ring with pure electric elements. In each case, simulations were done for particles with a zero EDM moment and for particles with an EDM moment of $\eta =10^{-15}$.}
	\label{SpinFrozen_perfect}
\end{figure}

\begin{figure}[tb]
	\centering
	\subfloat[]{\includegraphics[width=0.48\textwidth]{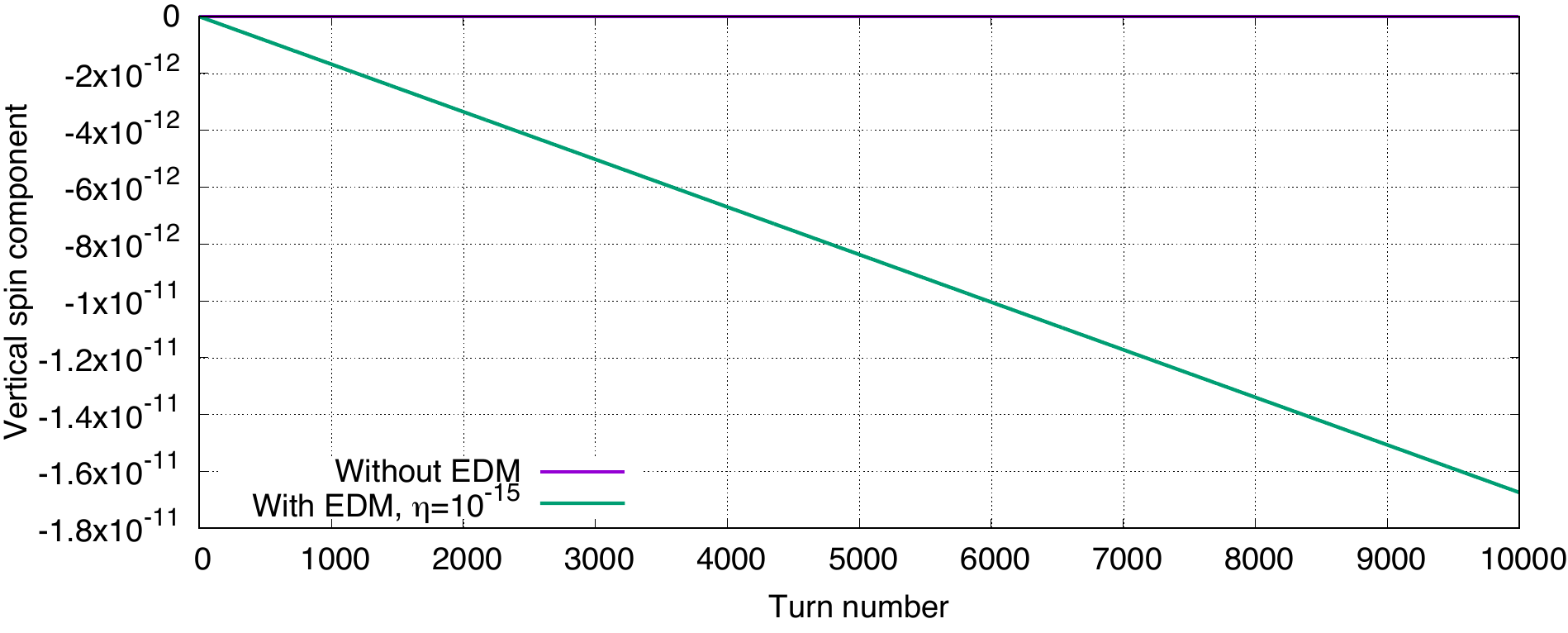}}\label{VerticalSpin_COx:dEDM}
	\subfloat[]{\includegraphics[width=0.48\textwidth]{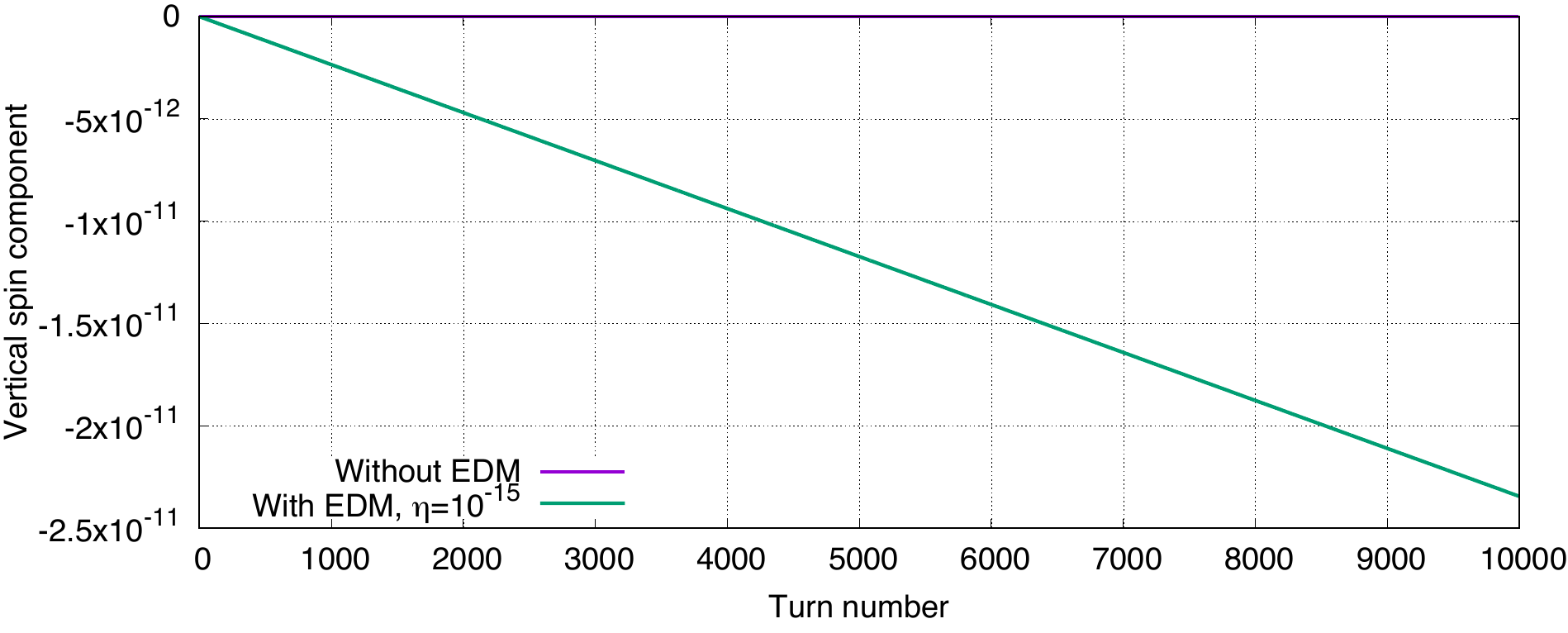}}\label{VerticalSpin_COx:pEDM}
	\caption{The vertical spin component as a function of turn number for a particle on the closed orbit where misalignments give a non-zero horizontal closed orbit but the vertical closed orbit is zero. a) Using the deuteron EDM ring. b) Using the proton EDM ring.}
	\label{verticalSpinBuildup_COx}
\end{figure}

\begin{figure}[tb]
	\subfloat[]{\includegraphics[width=0.48\textwidth]{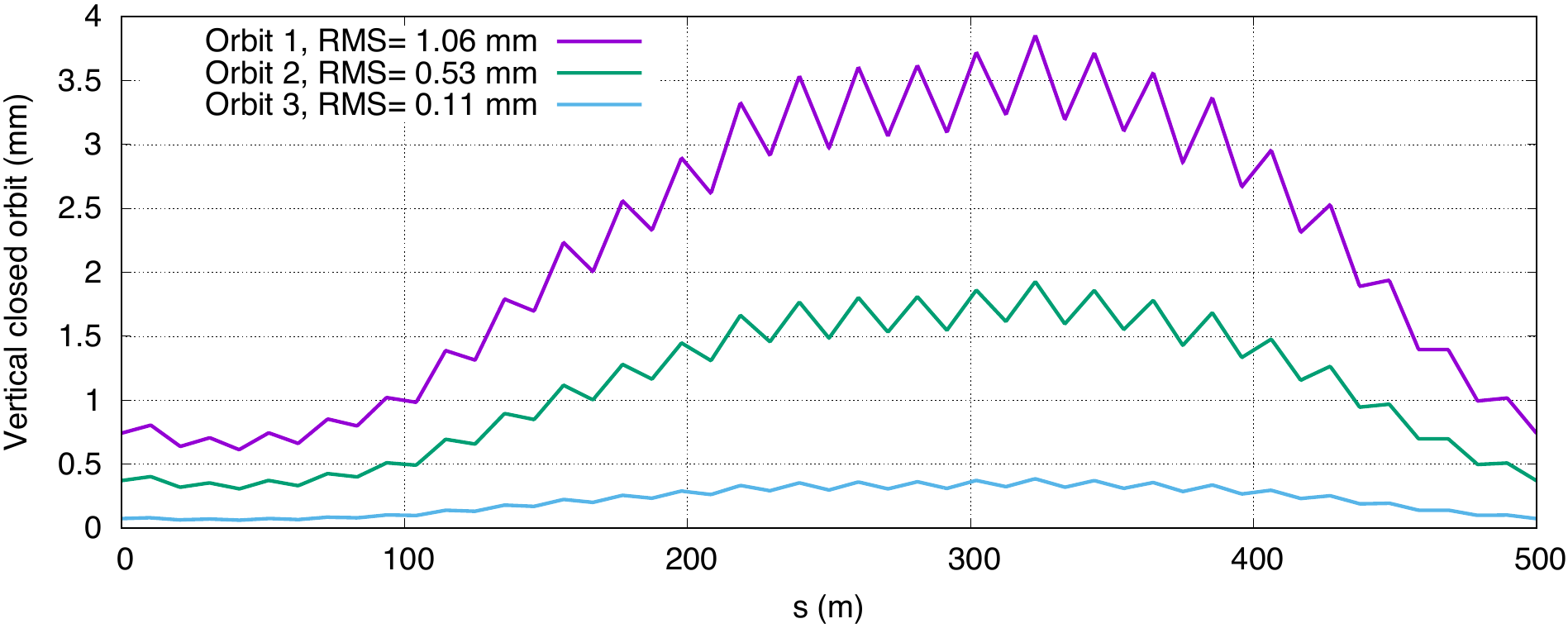}}
	\hfill
	\subfloat[]{\includegraphics[width=0.48\textwidth]{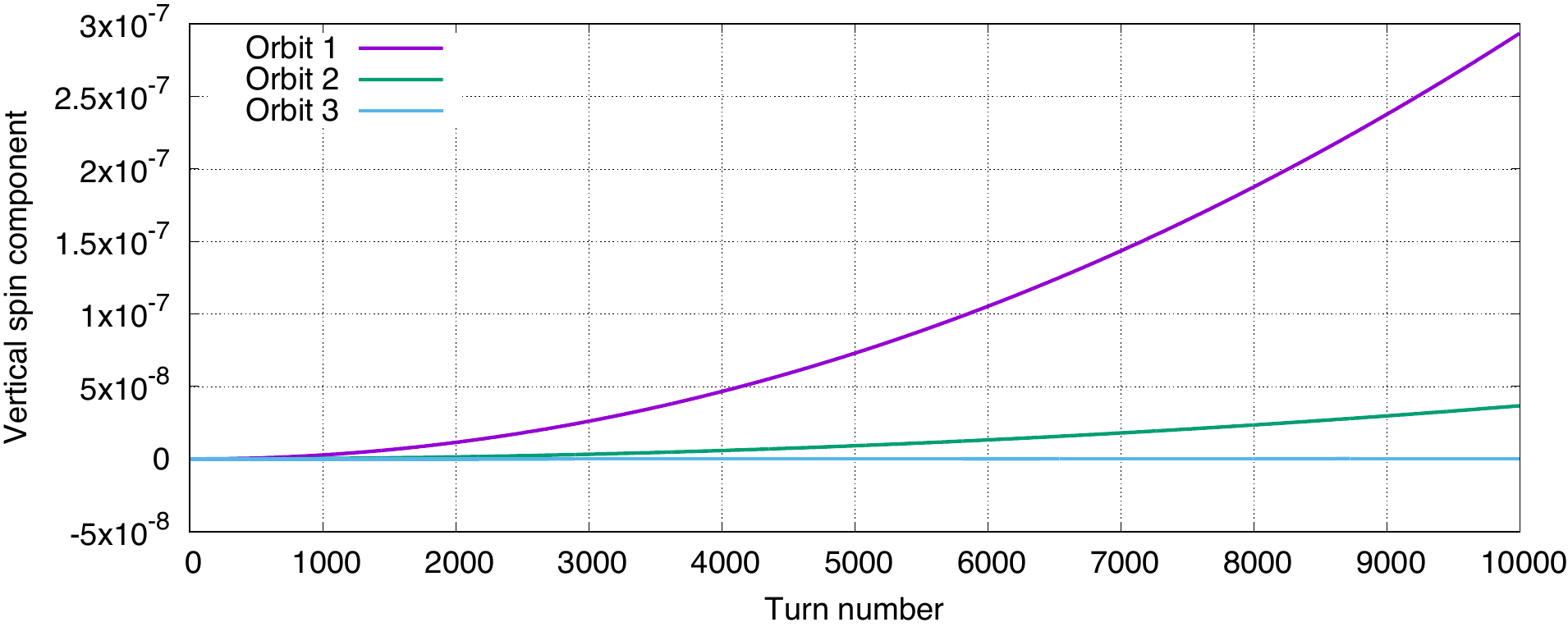}}
	\caption{a) Vertical closed orbits for three different quadrupole vertical misalignments in the deuteron ring. b) The corresponding vertical spin component buildup of a particle on the closed orbit. Here the EDM of the particle was set to zero.}
	\label{EB_verticalSpinBuildup_COy}
\end{figure}

\subsection{Preliminary Simulation Results}

To study the benefits of using MDMST quadrupoles, two lattices were employed. One was the lattice of Y. Semertzidis et al\cite{pEDM}. This is a proposed design for an EDM search using a pure electric proton storage ring. The other lattice is based upon the firest lattice but uses deuterons with MDMST quadrupoles and full spin frozen EB deflectors. Fig.~\ref{ringlayout} shows the layout of the deuteron ring, as well as the optics of this lattice.

The first study had no misalignments. That is, the closed orbit is through the center of all quadrupoles. Fig.~\ref{SpinFrozen_perfect} shows the average vertical component of the spin, as a function of orbital turn number, for 8 particles uniformly distributed with a normalized  vertical invariant $y \, y' \, \beta \, \gamma$ of $0.05\, \pi \, \text{mm-mrad}$. Fig.~\ref{SpinFrozen_perfect}a shows the simulation for the deuteron ring and Fig.~\ref{SpinFrozen_perfect}b shows the simulation for the proton ring. All simulations here are done using the Bmad toolkit\cite{b:bmad}. For each ring, the simulation was carried out for particles with and without an EDM moment. A value of $\eta =10^{-15}$ was used for the particles with a moment. For a deuteron, $\eta =10^{-15}$ corresponds to an EDM on the order of $10^{-29}$~e$\cdot$cm. Comparing the results with and without a finite EDM moment shows that the vertical spin buildup is due to the EDM part of the spin motion.

Fig.~\ref{verticalSpinBuildup_COx} shows a second study where the vertical spin component is plotted as a function of turn number for a particle launched on the closed orbit for both the deuteron and proton lattices. In this case, the quadrupoles in the lattices were misaligned to give a nonzero horizontal closed orbit but with a zero vertical closed orbit. It is evident that, in the absence of vertical motion the nonzero horizontal orbit has no impact on the vertical spin motion.

A vertical closed orbit distortion, on the other hand, can result in non-zero vertical spin buildup from the particle's magnetic dipole moment. This is illustrated in Fig.~\ref{EB_verticalSpinBuildup_COy} which shows the vertical spin component buildup of a particle with zero EDM in the deuteron storage ring with the quadrupoles vertically misaligned. Here, the magnetic dipole moment coupled with the vertical closed orbit leads to a buildup of the vertical spin component. The larger the vertical closed orbit, the faster the vertical spin component buildup gets.

\begin{figure}[tb]
	\subfloat[]{\includegraphics[width=0.48\textwidth]{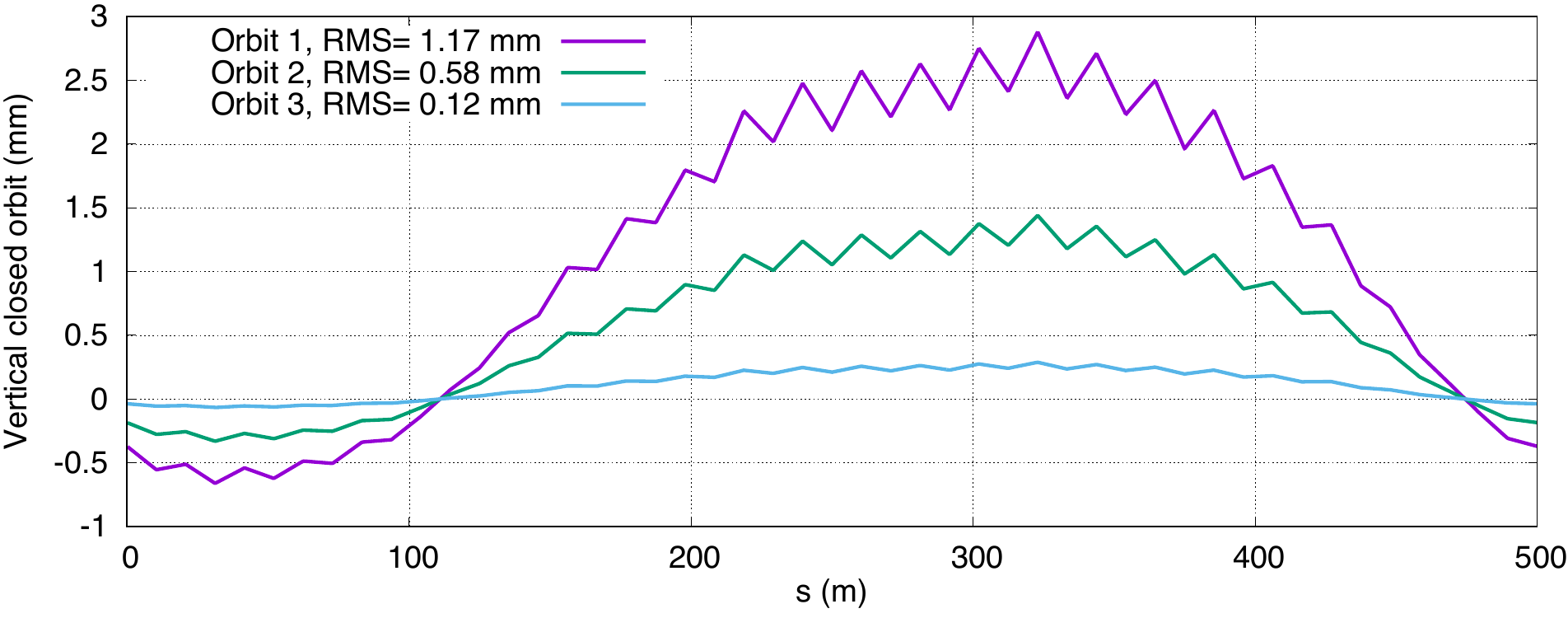}}
	\hfill
	\subfloat[]{\includegraphics[width=0.48\textwidth]{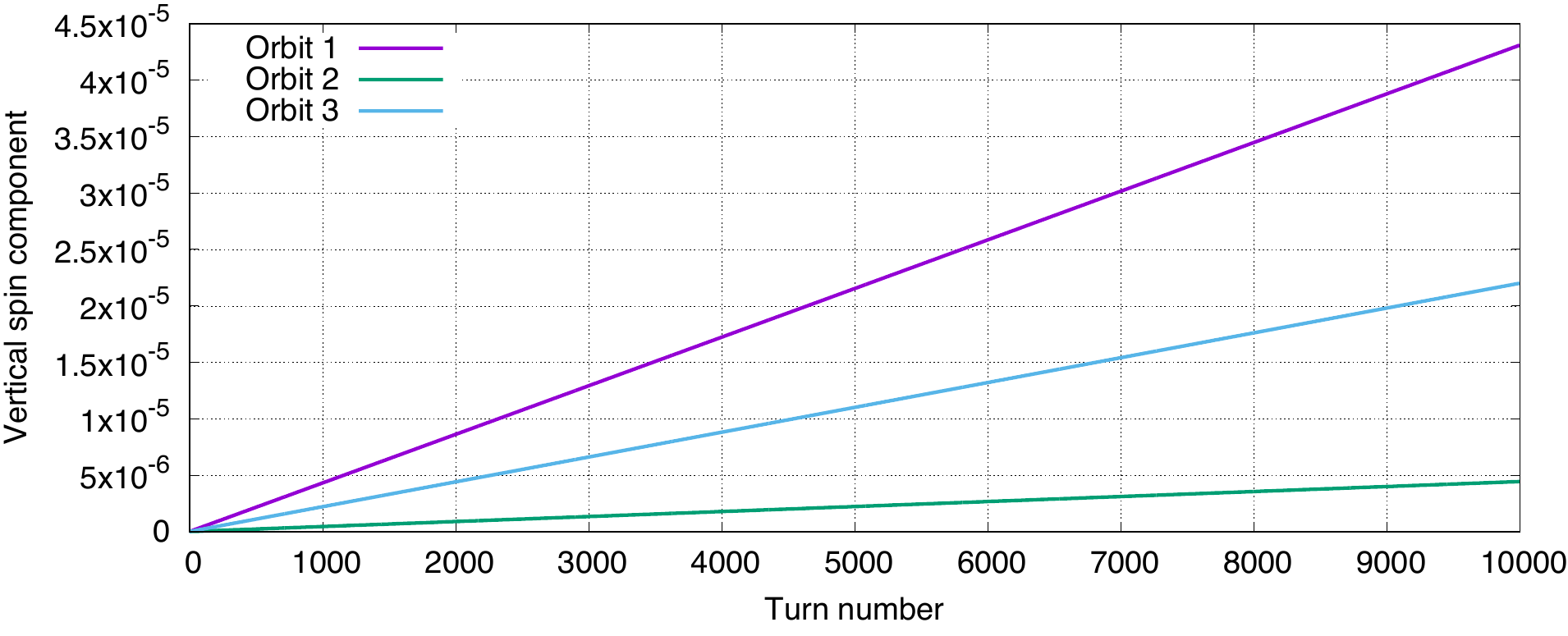}}
	\caption{a) Vertical closed orbits for three different quadrupole vertical misalignments in the proton ring. b) The corresponding vertical spin component buildup of a particle on the closed orbit. The proton EDM was set to zero.}
	\label{pEDM_verticalSpinBuildup_COy}
\end{figure}

 Similarly, the vertical spin component in the proton EDM storage ring with pure electric elements where the quadrupoles are misaligned also exhibits a linear buildup due to the magnetic dipole moment part of the spin motion, as shown in Fig.~\ref{pEDM_verticalSpinBuildup_COy}. 

This effect can be explained by the fact that the particle is going off-center of the quadrupole and its velocity is no longer perfectly perpendicular to the field it encounters. Assuming that the spin is perfectly frozen at each deflector, the one turn spin transfer map for a storage ring with thin lens quadrupoles is given by 
\begin{equation}
OTM=\prod_{i=1}^{N} M_i=e^{-\frac{i}{2}\psi\sigma_1}
\end{equation}
where $M_i=e^{-\frac{i}{2}\psi_i\sigma_1}$ is the spin transfer matrix of the $i^{th}$ quadrupole and the amount of spin precession $\psi_i$ is given by
\begin{equation}
\psi_i=(G\gamma+\frac{\gamma}{\gamma +1})\frac{\vec{E}_{y,i}\times \vec{\beta}}{c}\frac{L_i}{B\rho}
\end{equation}
with $L_i$ being the length of the quadrupole and $E_{y,i}$ is the electric field that the particle encounters at the $i^{th}$ quadrupole. For the case of a pure electric storage ring, with a non-zero vertical closed orbit, this then becomes
\begin{equation}
\psi_i=(G\gamma+\frac{\gamma}{\gamma +1})b_{1e,i}y_i\frac{\beta_{\parallel, i}}{c}\frac{L_i}{B\rho}
\end{equation}
where $\beta$ is the total velocity, $\beta_{y}$ is the velocity component along the vertical direction and $\beta^2=\beta^2_{\parallel}+\beta^2_{y}$. The total spin precession $\psi=\sum_{i=1}^{N}\psi_i$ after one orbital revolution is given by
\begin{equation}
\psi=(G\gamma+\frac{\gamma}{\gamma +1})\sum_{i=1}^{N} b_{1e,i}y_i\frac{\beta_{\parallel, i}}{c}\frac{L_i}{B\rho},
\label{Equad_psi}
\end{equation}
and is not equal to zero even though $\sum_{i=1}^{N} b_{1e,i}y_i L_i= 0$ for a stable particle on a non-zero closed orbit. Hence, the spin vector of the particle gets systematically lifted out of the horizontal plane and results in a non-zero vertical spin buildup.

\begin{figure}[tb]
	\centering
	\includegraphics[width=0.8\textwidth]{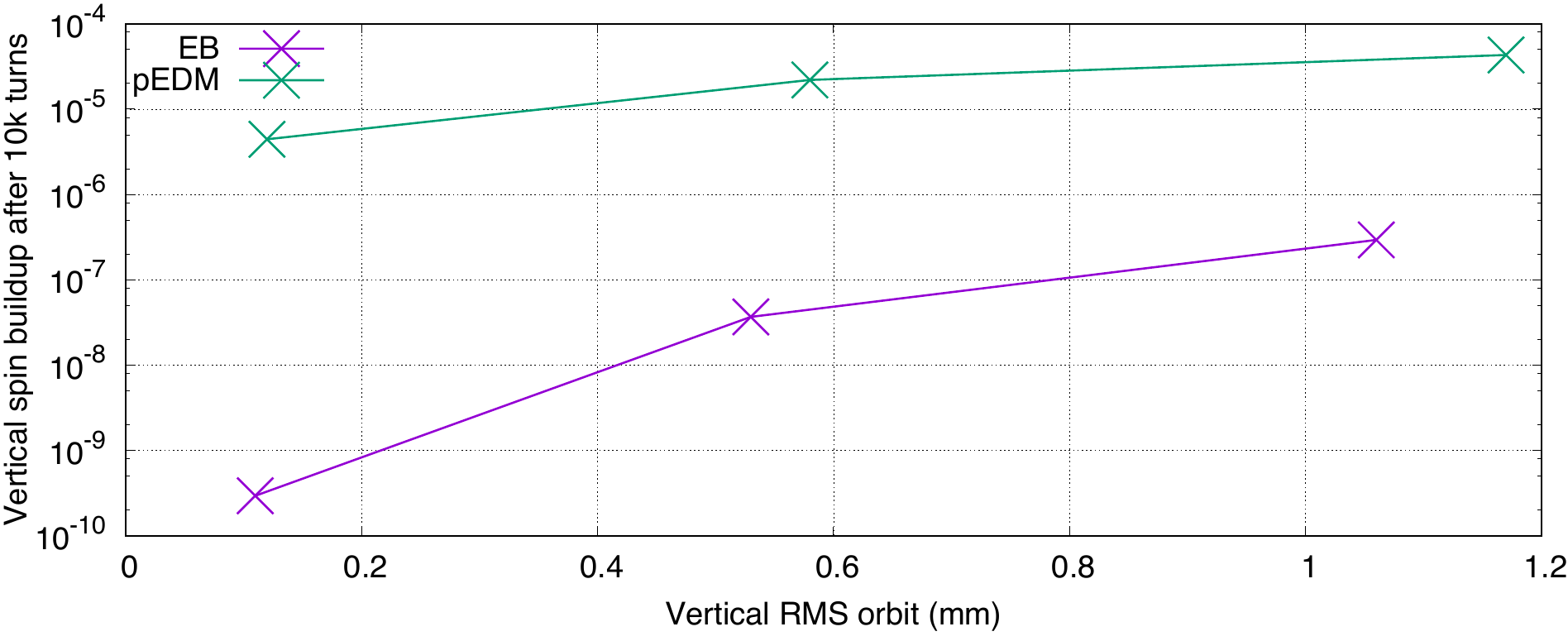}
	\caption{Vertical spin buildup rate as a function of the size of the vertical orbit distortion for the cases in Fig.~\ref{EB_verticalSpinBuildup_COy} and Fig.~\ref{pEDM_verticalSpinBuildup_COy} for the deuteron storage ring and the pure electric proton storage ring.}
	\label{both_builduprate_COy}
\end{figure}

For the case of the deuteron EDM ring with MDMST quadrupoles,  the amount of spin precession $\psi_i$ is given by
\begin{equation}
\psi_i=[(1+G\gamma)B_{x,i}+(G\gamma+\frac{\gamma}{\gamma +1})\frac{\vec{E}_{y,i}\times \vec{\beta}}{c}]\frac{L_i}{B\rho}
\label{EBquad_psi_i}
\end{equation}
where $B_{x,i}$ and $E_{y,i}$ are the magnetic and electric fields at the $i^{th}$ MDMST quadrupole, respectively. And, both satisfy the spin transparent condition given in Eq.~\ref{spintransparent_BE_relation}. That is
\begin{equation}
B_{x,i}=-(1-\frac{1}{(\gamma +1)(1+G\gamma)})\frac{\beta E_{y,i}}{c}
\end{equation}
Hence, the total amount of spin rotation $\psi$ in one orbital turn is
\begin{equation}
\psi=(G\gamma+\frac{\gamma}{\gamma +1})\sum_{i=1}^{N} b_{1e,i}y_i\frac{\beta}{c}(\frac{\beta^2_y}{\beta^2})\frac{L_i}{B\rho}.
\label{EBquad_psi}
\end{equation}
Since $\beta_y \ll \beta_{\parallel}$, it is evident that the vertical spin buildup rate in the deuteron storage ring with MDMST quadrupoles should be significantly smaller than it is in the pure electric storage ring. Fig.~\ref{both_builduprate_COy} shows the vertical spin buildup rate as a function of the size of the vertical orbit distortion for the cases in Fig.~\ref{EB_verticalSpinBuildup_COy} and Fig.~\ref{pEDM_verticalSpinBuildup_COy}. It is evident that overall the vertical spin buildup due to the MDM part of the spin motion is significantly smaller for the storage ring with MDMST quadrupoles, a clear advantage of this unique technique.

\begin{figure}[tb]
	\subfloat[]{\includegraphics[width=0.333\textwidth]{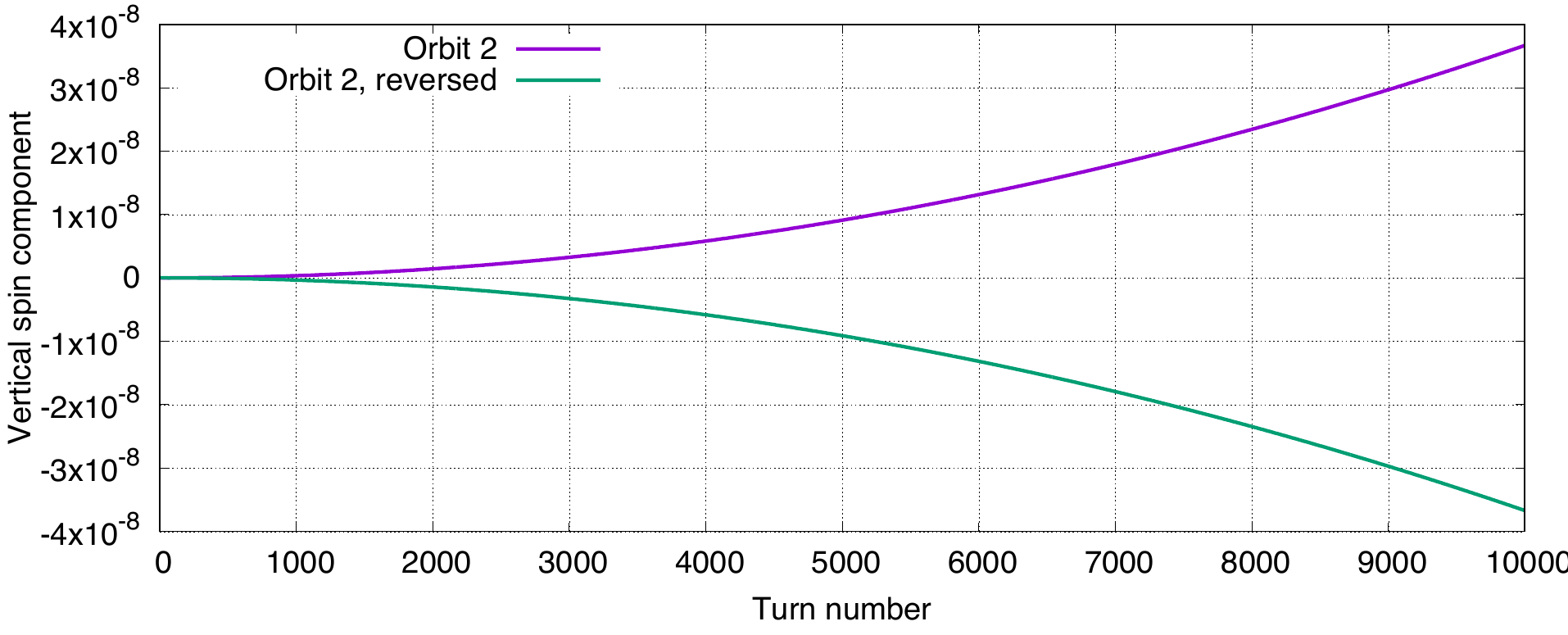}\label{EB_both_COy_sum:f1}}
	\hfill
        \subfloat[]{\includegraphics[width=0.333\textwidth]{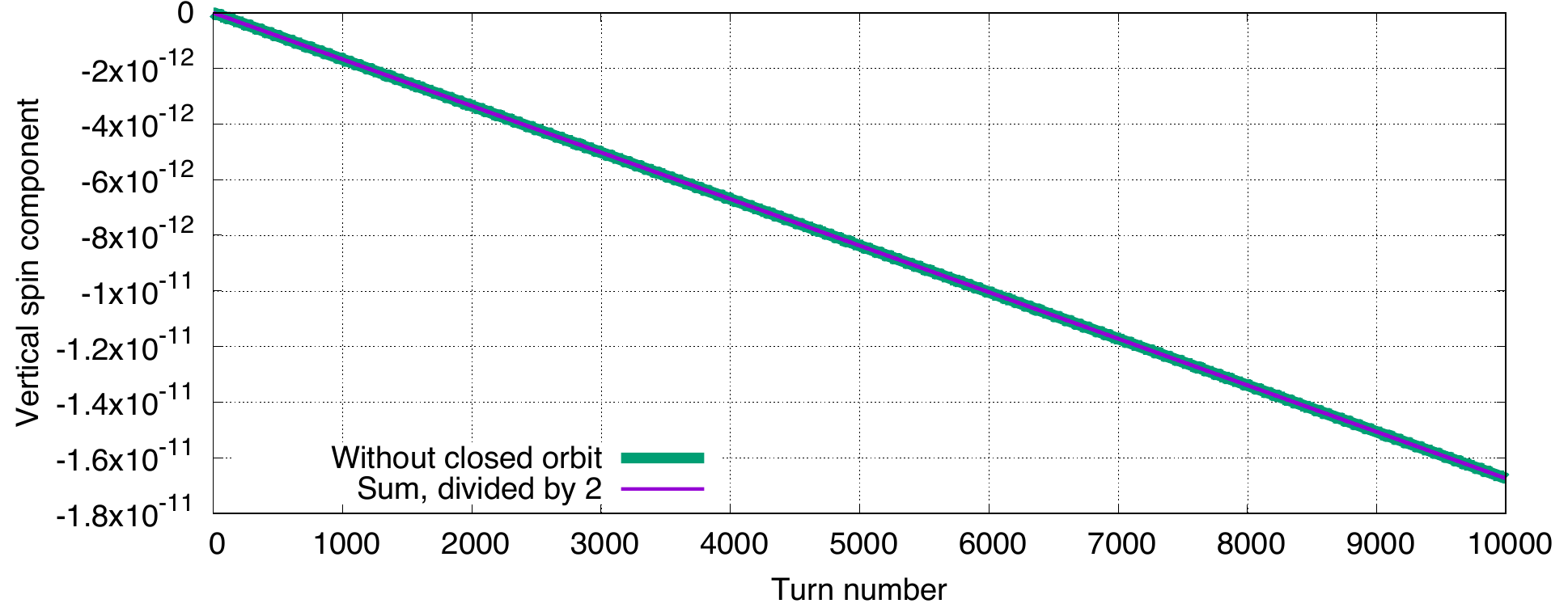}\label{EB_both_COy_sum:f2}}
        \hfill
        \subfloat[]{\includegraphics[width=0.333\textwidth]{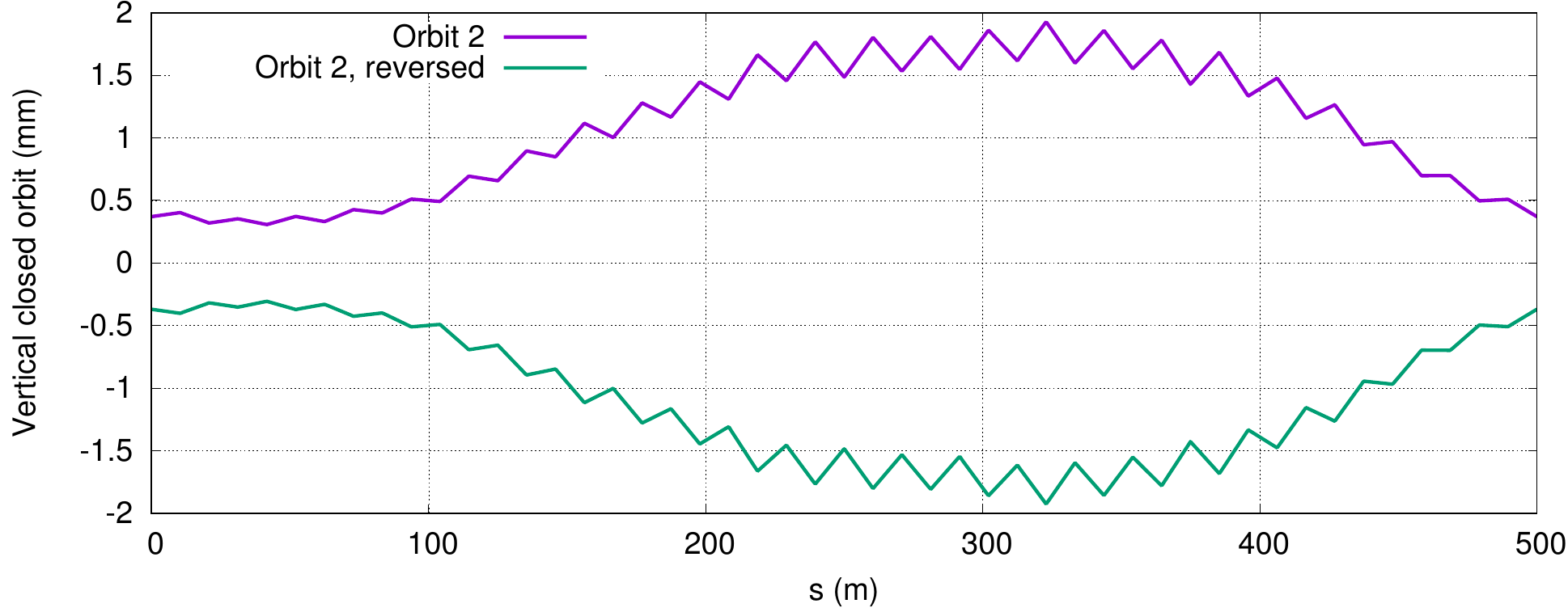}\label{EB_both_COy_sum:f3}}
	\caption{a) the vertical spin buildup in the deuteron ring with a vertical closed orbit that is reserved from Orbit 2 in Fig.~\ref{EB_verticalSpinBuildup_COy}. The vertical spin buildup corresponding to Orbit 2 in Fig.~\ref{EB_verticalSpinBuildup_COy} is also shown for comparison. b) The average vertical spin buildup of the two cases is virtually the same as the pure EDM driven vertical spin buildup in a perfect storage ring. c) The vertical orbit along the ring.}
	\label{EB_both_COy_sum}
\end{figure}

\begin{figure}[tb]
	\subfloat[]{\includegraphics[width=0.333\textwidth]{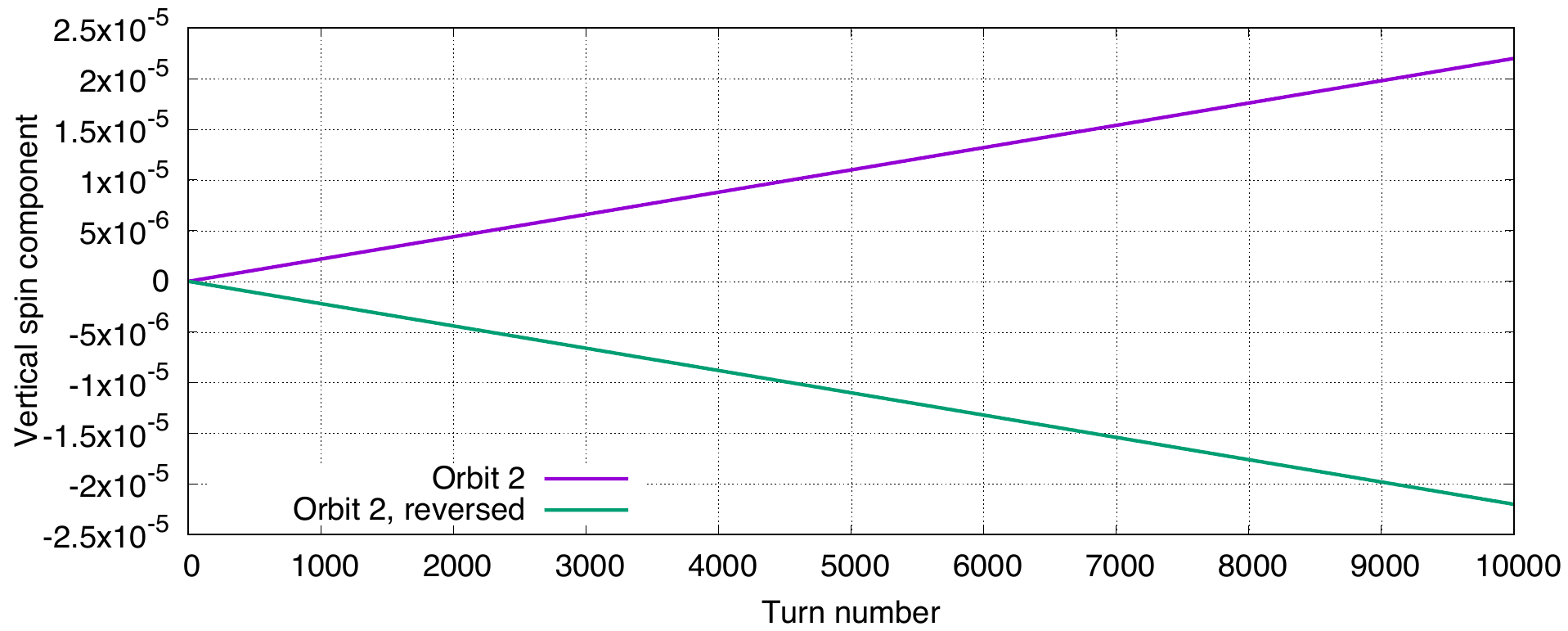}\label{pEDM_both_COy_sum:f1}}
	\hfill
	\subfloat[]{\includegraphics[width=0.333\textwidth]{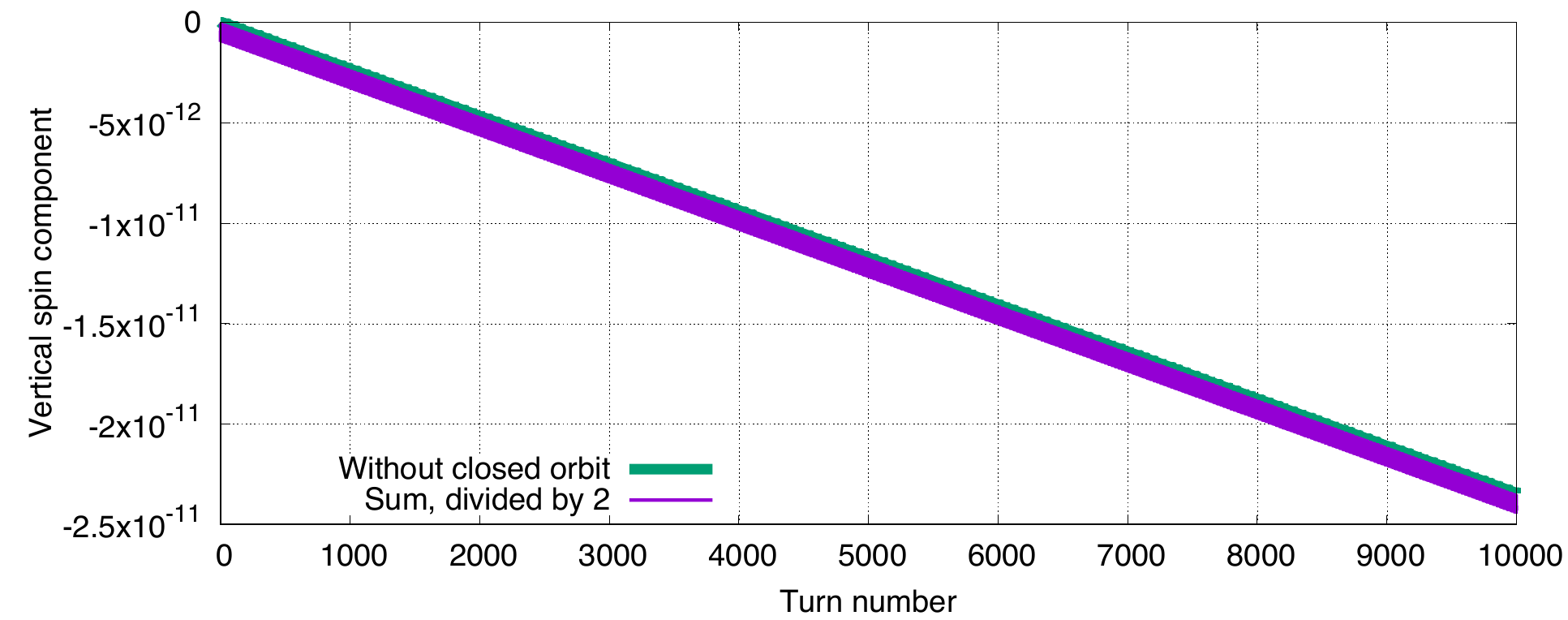}\label{pEDM_both_COy_sum:f2}}
        \hfill
        \subfloat[]{\includegraphics[width=0.333\textwidth]{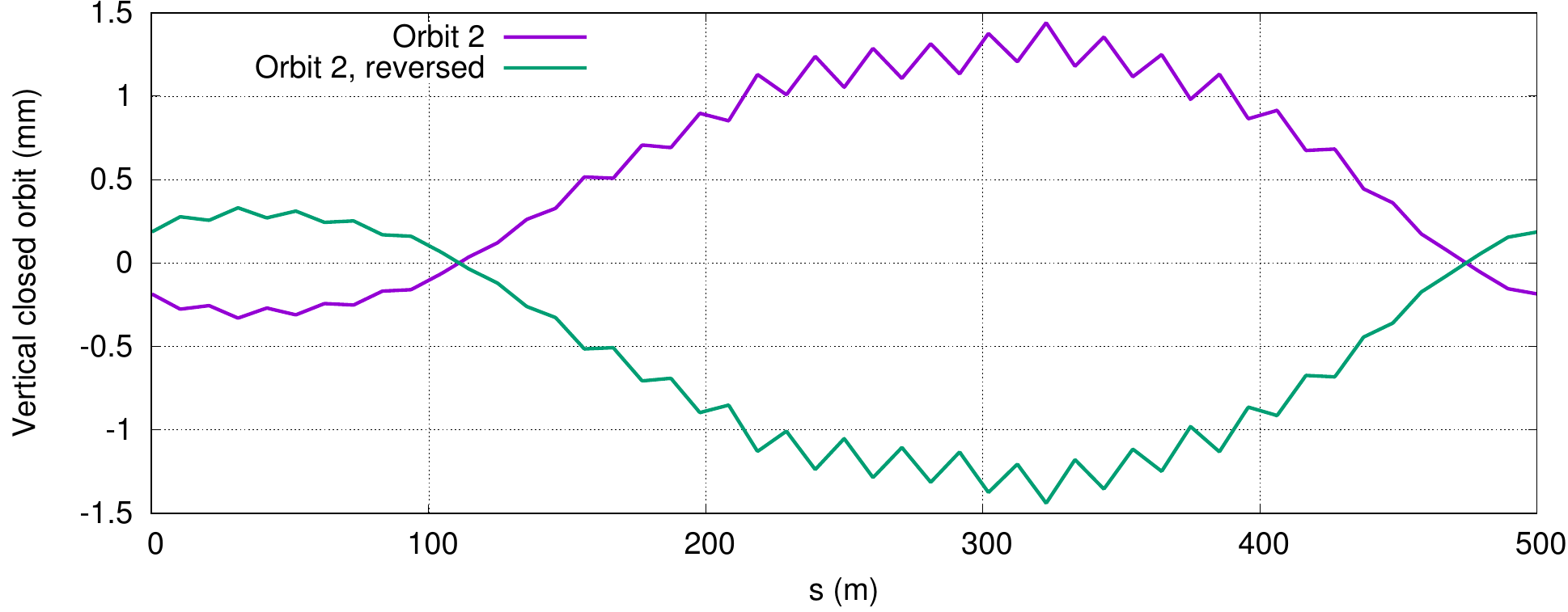}\label{pEDM_both_COy_sum:f3}}
	\caption{a) The vertical spin buildup in the proton ring with a vertical closed orbit that is reserved from the Orbit 2 in Fig.~\ref{pEDM_verticalSpinBuildup_COy}. The vertical spin buildup corresponding to Orbit 2 in Fig.~\ref{pEDM_verticalSpinBuildup_COy} is also shown for comparison. b) The average vertical spin buildup of the two cases is virtually the same as the pure EDM driven vertical spin buildup in a perfect storage ring. c) The vertical orbit along the ring.}
	\label{pEDM_both_COy_sum}
\end{figure}

In order to discriminate between the vertical spin component buildup due to the EDM verses the buildup due to the particle's MDM, we repeated the simulation for both the deuteron ring and the proton ring where we reversed the quadrupole misalignment that corresponds to the case of Orbit 2 in Fig.~\ref{EB_verticalSpinBuildup_COy} and Fig.~\ref{pEDM_verticalSpinBuildup_COy}. Fig.~\ref{EB_both_COy_sum:f1} and Fig.~\ref{pEDM_both_COy_sum:f1} confirm that the vertical spin buildup indeed reverses sign when the closed orbit is also reversed. Fig.~\ref{EB_both_COy_sum:f2} and Fig.~\ref{pEDM_both_COy_sum:f2} show that by averaging the vertical spin buildup, the EDM component of the vertical spin motion can be distinguished from the MDM component by either reversing the closed orbit or by reversing the direction of the beam. That is, using both clockwise~\(CW\) and counter clockwise~\(CCW\) beams provided that the orbit in both cases remains the same.

\section{Conclusion}

The concept of an MDM spin transparent quadrupole is proposed, and the feasibility of such an element for EDM storage ring is investigated. Our preliminary study including numerical simulation shows that a EDM storage ring with MDMST quadrupoles has the significant advantage of less systematic errors.  

Our next step is to carry out a set of studies to investigate the systematic error of the EDM storage ring with this unique element type. At the same time, we plan to investigate the design and feasibility of such an unique quadrupole.

\section{Acknowledgment}
The authors would like to thank Prof. A. Chao, Dr. E. Stephenson, Dr. W. Morse for their very inspiring and fruitful discussions and advise. The authors would like to thank Cornell University for its generous support on the development of Bmad, a very helpful and inclusive library for particle tracking including spin motion. In particular, the authors would like to thank Ms. Stockero from RWTH Aachen, who has been diligently helping scientists in the FZJ as well as in RWTH Aachen stay informed with various European funding opportunities. It was her effort that inspired us to think out of the box.


\begin{thebibliography}{66}
	\bibitem{srEDM}https://www.bnl.gov/edm/
	\bibitem{JEDI}http://collaborations.fz-juelich.de/ikp/jedi/about/introduction.shtml
	\bibitem{yannis} W. M. Morse et al, {\it rf Wien filter in an electric dipole moment storage ring: The "partially frozen spin" effect}, PRST-AB 16, 114001, 2013
	\bibitem{marcel} M. Rosenthal, http://collaborations.fz-juelich.de/ikp/jedi/private\_files/\\
	collaboration\_meeting/SystematicLimitations\_RF-WienFilterMethod.pdf, 2015 
	\bibitem{dEDM_BNL} Y. Semertzidis, W. M. Morse et al,\\ https://www.bnl.gov/edm/files/pdf/deuteron\_proposal\_080423\_final.pdf, 2008
	\bibitem{pEDM_BNL} Y. Semertzidis, W. M. Morse et al,\\
	https://www.bnl.gov/edm/files/pdf/Proton\_EDM\_proposal\_20111027\_final.pdf, 2011
	\bibitem{Ed} E. Stephenson, http://collaborations.fz-juelich.de/ikp/jedi/private\_files/\\
	collaboration\_meeting/Stephenson-RingsPolarimeters.pdf
    \bibitem{pEDM} V. Anaslassopoulos et al, arXiv:1502.04317, 2015
    \bibitem{b:bmad} D. Sagan, ``Bmad: A relativistic charged particle simulation,''
Nuc.\ Instrum.\ Methods Phys.\ Res.\ A, {\bf 558}, pp 356-59 (2006).
\end{thebibliography}
\end{document}